\documentclass[aps,prl,letterpaper,twocolumn,floatfix,showpacs,amsmath]{revtex4}
\usepackage{graphicx,amsmath}

\newcommand{\Eq}[2][Eq.~]{{#1}(\ref{eq:#2})}
\newcommand{\Eqs}[1]{Eqs.~(\ref{eq:#1})}
\newcommand{\Fig}[2][Fig.~]{{#1}\ref{fig:#2}}

\newcommand{\Df}{{D_{\hspace{-1pt}f}}}

\renewcommand{\Re}{\mathrm{Re}\,}

\newcommand{\qquadtext}[1]{\qquad\text{#1}\qquad}

\newcommand{\ld}{\lambda}

\renewcommand{\th}{\theta}
\newcommand{\ens}[2][]{\langle{#2}\rangle_{#1}}

\begin{document}

\title{Diffusion-Limited Aggregation on Curved Surfaces}

\author{Jaehyuk Choi$^1$, Darren Crowdy$^2$ and Martin Z. Bazant$^{1,3}$ }
\affiliation{$^1$ Department of Mathematics, 
  Massachusetts Institute of Technology, Cambridge, MA 02139}
\affiliation{$^2$ Department of Mathematics, Imperial College, London, UK}
\affiliation{$^3$ Department of Chemical Engineering, Massachusetts Institute of Technology, Cambridge, MA 02139 }

\date{\today}

\begin{abstract}
We develop a general theory of transport-limited aggregation phenomena occurring on curved surfaces, 
based on stochastic iterated conformal maps and conformal projections to the complex plane. To illustrate the theory, we use stereographic projections to simulate diffusion-limited-aggregation (DLA) on surfaces of constant Gaussian curvature, including the sphere ($K>0$) and pseudo-sphere ($K<0$), which approximate ``bumps" and ``saddles" in smooth surfaces, respectively.  Although curvature affects the global morphology of the aggregates,
the fractal dimension (in the curved metric) is remarkably insensitive
to curvature, as long as the particle size is much smaller than the
radius of curvature. We conjecture that all aggregates grown by conformally invariant transport on curved surfaces have the same fractal dimension as DLA in the plane.  Our simulations suggest, however, that the multifractal dimensions increase from hyperbolic ($K<0$) to elliptic ($K>0$) geometry, which we attribute to curvature-dependent screening of tip branching.
\end{abstract}


\maketitle


The Laplacian growth model and its stochastic analogue,
diffusion-limited aggregation (DLA)~\cite{witten81}, describe the essential physics of non-equilibrium pattern formation in diverse situations~\cite{bunde},
such as viscous fingering~\cite{bensimon86}, 
(quasi-static) dendrite solidification~\cite{kessler88}, 
dielectric breakdown~\cite{niemeyer84}, and dissolution~\cite{TLD}, depending on conditions at the moving, free boundary. Some extensions to non-Laplacian growth phenomena, such as advection-diffusion-limited aggregation~\cite{adla,david05} (ADLA) and brittle fracture~\cite{barra02}, are also available. 
Almost all prior work with these models has assumed flat Euclidean space, typically a two-dimensional plane, 
but real aggregates, such as mineral dendrites~\cite{chopard91}, cell colonies~\cite{wang97},  and cancerous tumors~\cite{ho98}, often grow on curved or rough surfaces.
Aside from a few studies of Eden-like clusters~\cite{wang97} and continuous viscous 
fingers~\cite{entov97, parisio01}  on spheres, it seems that the effects of surface curvature on pattern formation have not been investigated. 

In this Letter, we extend transport-limited growth models to curved two-dimensional surfaces via  
conformal projections from the plane.  Time-dependent conformal maps are widely used in physics~\cite{gruzberg04} and materials science~\cite{handbook} to describe interfacial dynamics in two dimensions. Continuous conformal maps have long been applied to  viscous fingering~\cite{howison92,bensimon86}, and more recently,  Hastings and Levitov introduced stochastic, iterated conformal maps for DLA~\cite{hastings98}. Both continuous and stochastic conformal-map dynamics have also been extended to other conformally invariant (but non-Laplacian and nonlinear) gradient-driven transport processes~\cite{bazant04,adla,david05}, such as advection-diffusion in a potential flow~\cite{kornev94,choi05} or electrochemical transport in a quasi-neutral solution~\cite{bazant04}.  Indeed, there is nothing special about harmonic functions (solutions to Laplace's equation) in the plane, aside from the direct connection to analytic functions of  a complex variable (real or imaginary part). The key property of conformal invariance is shared by other equations~\cite{bazant04,handbook} and, as note here, applies equally well to conformal (i.e. angle preserving) transformations between curved surfaces. Here, we formulate continuous and discrete conformal-map dynamics for surfaces of constant curvature, by a sequence of mappings from the complex plane, and we use the approach to study the fractal and multi-fractal properties of DLA on curved surfaces.

\begin{figure} \centering
  \includegraphics[width=0.8\linewidth]{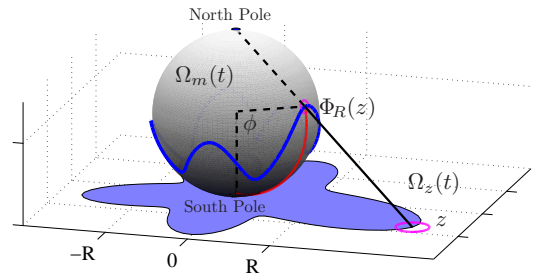}
  \caption[Stereographic projection]{ 
    \label{fig:shadow}
    Stereographic projection, $\Phi^{-1}$, from the exterior of the 
    growing object $\Omega_m(t)$ on a sphere of radius $R$ to the exterior
    of the shadow, $\Omega_z(t)$, on a complex plane.
    The point $\Phi(z)$ is projected from the north pole to the
    point $z$. The origin of the $z$-plane is tangent to the sphere 
    at the south pole, and the latitudinal angle $\phi$ is measured
    from the south pole.
  }
\end{figure}

\begin{figure} \centering
  \raisebox{-\height}{(a)}
  \raisebox{-\height}{
    \includegraphics[width=0.4\linewidth]{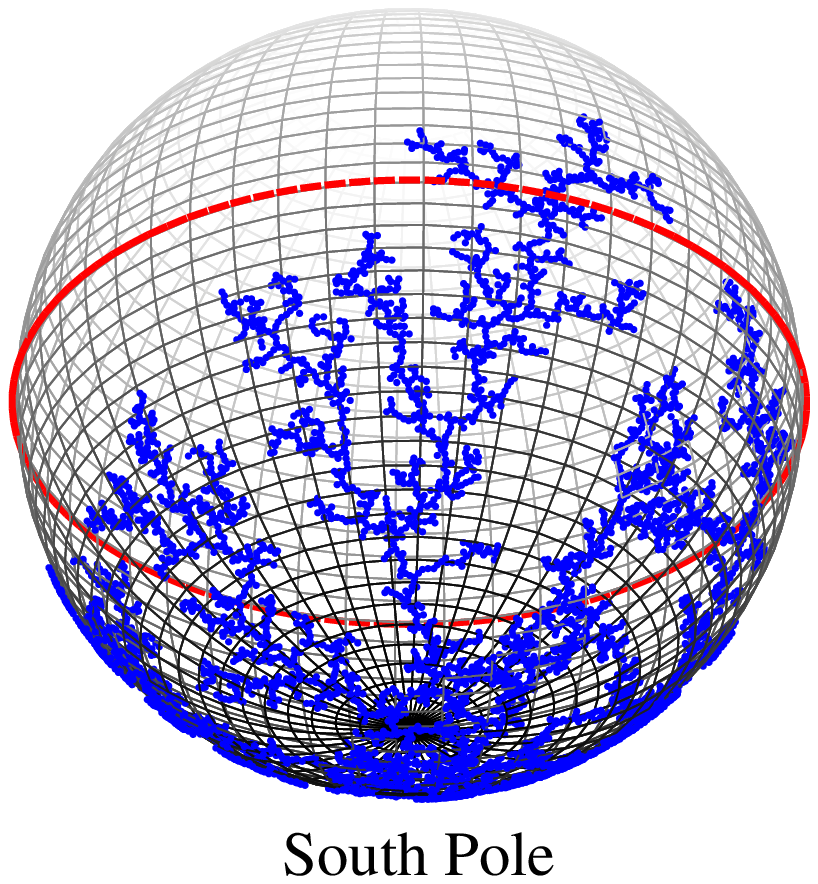}
  }\;
  \raisebox{-\height}{(b)}
  \raisebox{-\height}{
    \includegraphics[width=0.4\linewidth]{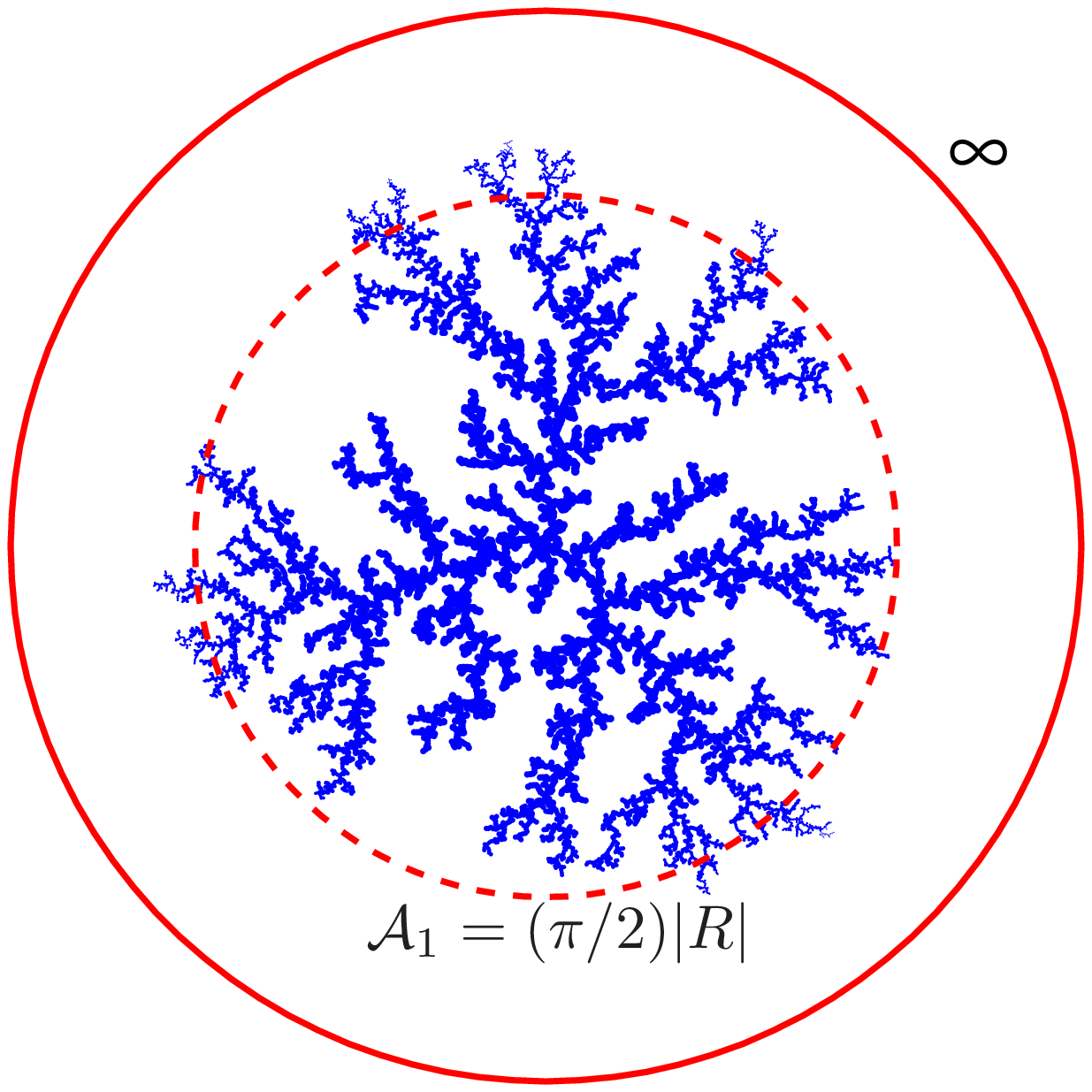}
  }
  \caption[DLA clusters on the elliptic and the hyperbolic geometries]{ 
    \label{fig:cluster}
    DLA clusters on the elliptic (left) and the hyperbolic (right) geometries.
    The elliptic geometry is isometrically embedded on the surface of a 
    sphere and the hyperbolic geometry is visualized on the Poincar\'e 
    disk with the metric, $ds = dz/(1-(|z|/2R)^2)$. 
    We use $R/\sqrt{\ld_o} = 100$ and aggregate
    4942 and 9001 particles to fill the great circles of 
    radius $(\pi/2) R$ (dashed lines) in the elliptic and hyperbolic
    geometries respectively.
  }
\end{figure}

{\it Transport-limited growth on curved surfaces. ---}
At first, it would seem that conformal-map dynamics cannot be directly applied to a non-Euclidean geometry, and certainly the formulation could not be based only upon analytic functions of the complex plane. Nevertheless, if there exists a conformal (angle-preserving) map between the curved manifold and the plane, and if the underlying transport process is conformally invariant~\cite{bazant04,choi05,adla,david05,handbook}, then any solution to the transport equations in the non-Euclidean geometry can be conformally mapped to a solution in the complex plane, without changing its functional form. Such maps do exist, as we illustrate below with important special cases.
 
Let $\Omega_m(t)$ be the exterior of a growing object on a non-Euclidean
manifold, $M$, and $\Phi$ be a conformal map from a (part of) complex plane
to $M$. We speak of the domain $\Omega_m(t)$ as having its {\it shadow}, 
$\Omega_z(t) = \Phi^{-1}(\Omega_m(t))$, on the complex plane 
under the {\it projection} $\Phi^{-1}$.
As in the flat surface case, we can describe the growth by a time-dependent 
conformal map, $g(w,t)$, from the exterior of the unit disk, $\Omega_w$, 
to the exterior of the growing shadow, $\Omega_z(t)$, which in turn is mapped onto the curved geometry by the inverse projection. Care must only be taken that the dynamics of $g(w,t)$ should describe $\Omega_z(t)$ in such a way that the evolving object, 
$\Omega_m(t)$, follows the correct physics of growth on $M$, rather than in the intermediate complex plane, which is purely  a mathematical construct.

Let us illustrate this point for both continuous and discrete versions of conformally invariant, transport-limited growth~\cite{adla,handbook}.
For continuous growth, we generalize the Polubarinova-Galin equation
equation~\cite{polub45, adla,handbook} for a curved manifold with a conformal
map $\Phi(z)$ as follows
\begin{equation}
  \label{eq:polub}
  \Re\{\overline{w\  g'(w)}\ g_t(w)\} = \frac{ \alpha\; \sigma(w,\, t) }{ 
  |\Phi'\circ g(w)|^2 } .
\end{equation}
for $|w|=1$, where $\alpha$ is a constant and $\sigma(w,\,t)$ is the
time-dependent flux density on the boundary of $\Omega_w$.
This result is easily obtained by substituting $\Phi\circ g$ for $g$ in the
original equation. 

For stochastic growth, we adjust the 
Hastings-Levitov algorithm~\cite{hastings98} on the shadow domain. 
The algorithm is based on the recursive updates of the map,
\begin{equation}
  g_n(w) =  g_{n-1}\circ\phi_{\ld_n,\th_n}(w), \quad g_n(w) = g(w,t_n),
\end{equation}
where $\phi_{\ld,\th}$ is a specific map that slightly distort $\Omega_w$
by a bump of area $\ld$ around the angle $\th$.
While the random sequence $\{\th_n\}$ follows the probability distribution,
$p(\th,t_n) \propto \sigma(e^{i\th},t_n)$, invariant under conformal maps,
the preimage of bump area, $\ld_n$, should be determined so that the 
bump area is fixed as $\ld_0$ on the manifold, $M$.
Thus the bump size of the $n$-th accretion is determined by
\begin{equation}
  \label{eq:ld}
  \ld_n = \frac{ \ld_0 }{  |\Phi'\circ g_{n-1}(e^{i\th_n})|^2\cdot |g_{n-1}'(e^{i\th_n})|^2 } .
\end{equation}
We are not aware of any prior modification of the Hastings-Levitov algorithm with this general form. 
Previous studies on DLA 
in a channel geometry~\cite{somfai03a}
can be viewed as an example with $M=\{z:\; 0 < {\rm arg}\; z < 2\pi\}$ and  
$\Phi(z) = \log(z)$ although the manifold is Euclidean.

{\it Stereographic projections. }
To illustrate the general theory, we first make use of classical stereographic projection~\cite{needham} to describe growth on a sphere. 
Stereographic projection is obtained by projecting the surface of a sphere
from the north pole to a plane whose origin is tangent to the south 
pole; see \Fig{shadow}. If $\Phi$ is an {\it inverse} stereographic 
projection with sphere of radius $R$, 
$\Phi^{-1}$ maps the point $(R,\phi,\th)$ in spherical
coordinates to $z=R\tan(\phi/2)e^{i\th}$ in the complex plane. Here $\th$
is the azimuthal angle and $\phi$ is the latitudinal angle measured from
the south pole. If the modulus, $|\,\cdot\,|$, on the sphere is defined 
to be distance to the origin (south pole) in the curved metric
(arc-length of the great circle) in a similar way to $|\,\cdot\,|$ on a 
complex plane, $|z|$ and $|\Phi(z)|$ satisfy
\begin{equation}
  \label{eq:distance}
  \frac{|z|}{2R} = \tan\left( \frac{|\Phi(z)|}{2R} \right).
\end{equation}
The Jacobian factor of the projection is angle-independent; thus, it is given 
by
\begin{equation}
  \label{eq:jacobian}
  |\Phi'(z)| = \frac{d|\Phi(z)|}{d|z|} = \frac1{1 + (|z|/2R)^2}
\end{equation}
from the derivative of \Eq{distance}. Now \Eq{jacobian} can be 
used to continuous dynamics, \Eq{polub}, and stochastic dynamics, 
\Eq{ld} on sphere.

While the surface of a sphere is a three-dimensional visualization of 
elliptic (or Riemannian) geometry, we can also obtain a conformal projection
from hyperbolic geometry to a complex plane as well.
Unlike elliptic geometry, hyperbolic geometry can not be isometrically
embedded into 3D Euclidean space; only a part of the geometry can be embedded 
into 3D as a surface known as {\it pseudosphere}. 
Hyperbolic geometry has a negative constant curvature, $K = -1/R^2$, 
as opposed to the positive one, $K=1/R^2$, of elliptic geometry.
The projection can be obtained by simply viewing hyperbolic geometry as a
surface of a sphere with an imaginary radius, $iR$, as suggested from the 
sign of curvature. The projection can be still defined; substituting 
$iR$ for $R$ alters \Eqs{distance} and \Eq[]{jacobian} as
\begin{equation}
  \frac{|z|}{2R} = \tanh\left( \frac{|\Phi(z)|}{2R} \right) 
  \qquadtext{and} |\Phi'(z)| = \frac1{1 - (|z|/2R)^2}.
\end{equation}
The image of hyperbolic geometry under the projection is thus limited to the
inside of a disk with radius $2R$, and the boundary, $|z|=2R$, corresponds
to the infinity. The stereographic projection serves as a non-isometric 
visualization of the geometry known as Poincar\'e disk. 
The length element $dz$ at $dz$ in the disk corresponds to the element 
$dz/(1-(|z|/2R)^2)$ in the hyperbolic universe.

\begin{table} \centering
  \begin{tabular}{cccc} \\ \hline\hline
    Geometry & Elliptic & Euclidean & Hyperbolic \\ \hline
    $\Df$ & 1.704$\pm$0.001 & 1.704$\pm$0.001 & 1.693$\pm$0.001 \\ \hline
    $D_2/\Df$ & 0.577$\pm$0.003 & 0.529$\pm$0.006 & 0.527$\pm$0.001 \\
    $D_3/\Df$ & 0.591$\pm$0.006 & 0.503$\pm$0.005 & 0.499$\pm$0.001 \\
    $D_4/\Df$ & 0.617$\pm$0.012 & 0.486$\pm$0.004 & 0.482$\pm$0.001 \\
    $D_5/\Df$ & 0.631$\pm$0.018 & 0.473$\pm$0.004 & 0.469$\pm$0.001 \\ \hline\hline
  \end{tabular}
  \caption[The fractal dimension and the multifractal dimensions of DLA
    clusters on curved geometries]{ 
    \label{tb:dim}
    The fractal dimension $\Df$ and the multifractal dimensions $D_{2q+1}$ of DLA 
    clusters on three different geometries.
  }
\end{table}

\begin{figure} \centering
  \raisebox{-\height}{(a)}
  \raisebox{-\height}{
    \includegraphics[width=0.9\linewidth]{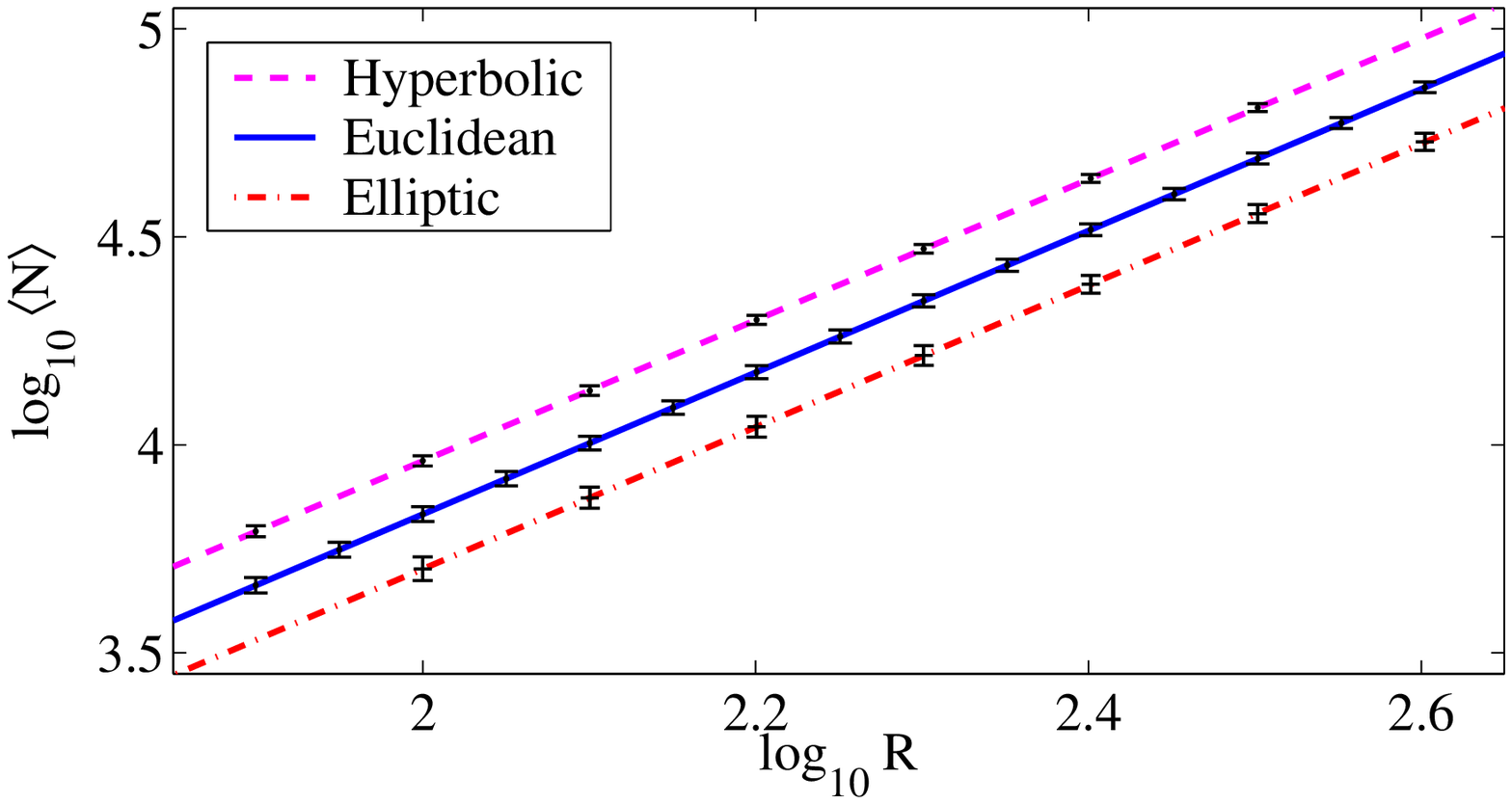}
  } \\ [1ex]
  \raisebox{-\height}{(b)}
  \raisebox{-\height}{
    \includegraphics[width=0.9\linewidth]{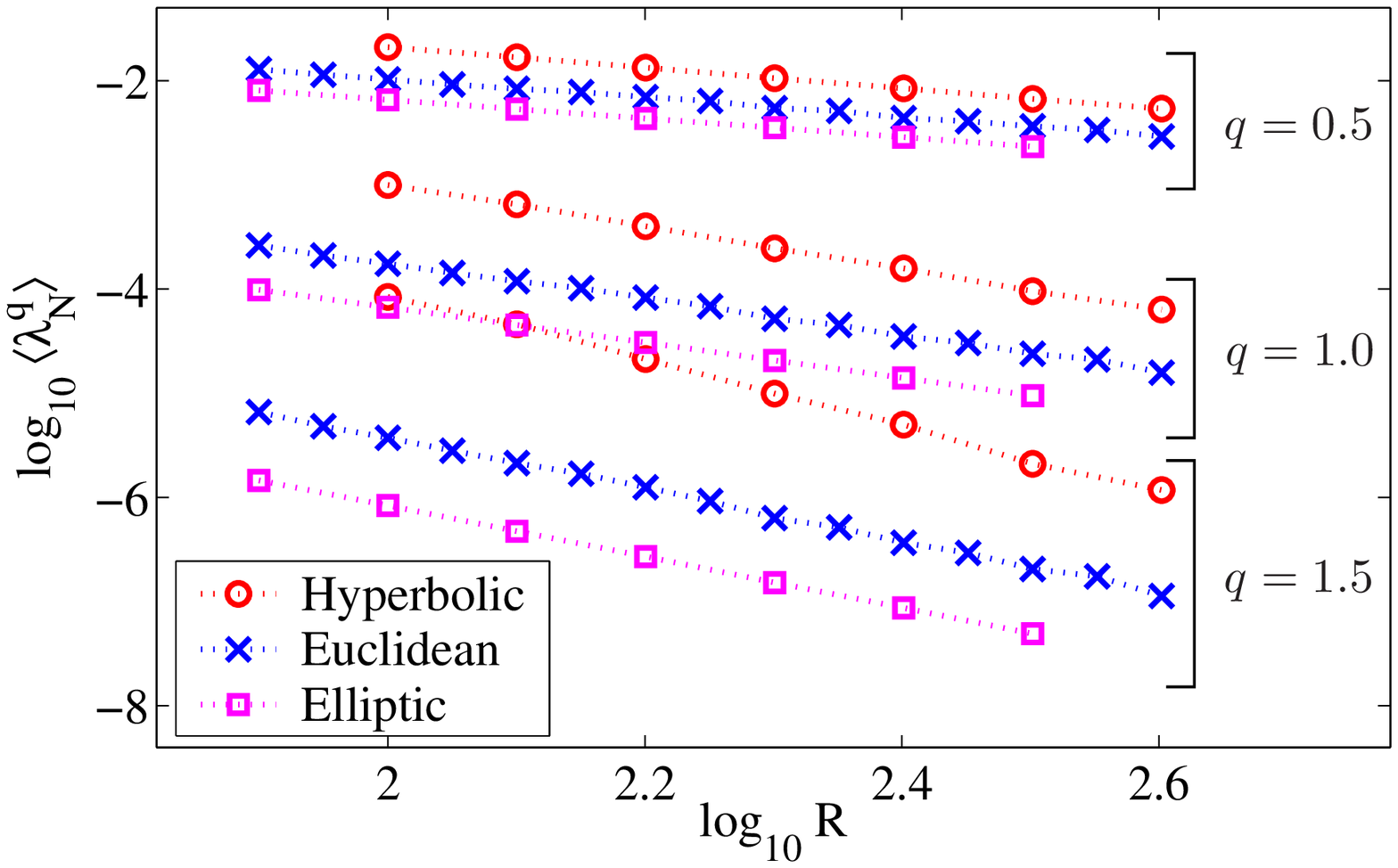}
  }
  \caption[The scalings of DLA clusters on constant curvature surfaces]{
    \label{fig:fracdim}
    (a) The average number of particles, $\ens{N}$, to fill the disk of conformal radius,
    $\mathcal{A}_1 = (\pi/2)R$, versus radius $R$
    on the three geometries.
    (b) Moments  $\ens{\ld_N^q}$ of the pre-image bump size, $\ld_N$, versus $R$, which define $D_{2q+1}$ via (\ref{eq:Dq}.
  }
\end{figure}

{\it  DLA on constant-curvature surfaces. ---}
Using the harmonic probability measure, $p(\th)=1/2\pi$, and the modified
Hastings-Levitov algorithm with \Eq{ld}, we grow DLA clusters on curved 
surfaces.
\Fig{cluster} shows clusters on elliptic and hyperbolic geometries. 
On the Poincar\'e disk, the size of particles becomes smaller and
smaller as they approach the infinity, $|z| = 2R$.

An analytic advantage of the conformal mapping formulation is 
that the Laurent expansion of $g(w,t)$ gives us information on
the moments of the cluster; the {\it conformal radius}, $A_1$, and 
the {\it center of charge}, $A_0$, come from the first two terms of
the expansion, $g(w) \approx A_1w + A_0$.
Such coefficients from DLA clusters on curved surfaces are not useful 
as they are the moments of the shadow, not the original cluster. 
Using the property that circles are mapped to circles under an (inverse) 
stereographic projection, it is possible to define analogous quantities corresponding to 
$A_1$ and $A_0$.
We note that the image, $\Phi(A_1w + A_0)$, of the unit circle, $|w|=1$,
is also a circle on $M$, and it is the one that best approximate
the cluster and the far-field in $\Omega_m(t)$.
Thus, we define the conformal radius, $\mathcal{A}_1$, and the center of 
charge, $\mathcal{A}_0$, on $M$ to be the radius and the deviation of 
$\Phi(A_1w + A_0)$ respectively:
\begin{gather}
  \label{eq:A1A0}
  \mathcal{A}_1 = \frac12\; \{ \; |\Phi(A_1+|A_0|)| + |\Phi(A_1-|A_0|)|\; \},\\
  \mathcal{A}_0 = \frac12\; \{ \; |\Phi(A_1+|A_0|)| - |\Phi(A_1-|A_0|)|\; \}.
\end{gather}
On curved surfaces, however, the fractal dimension $\Df$, cannot be determined
from the scaling, $n\sim\ens{\mathcal{A}_1}^\Df$, since the geometry is not 
linearly scalable with $\mathcal{A}_1$;
the two domains within radii of different values of $\mathcal{A}_1$ are not 
self-similar to each other, and the log-log plot of $n$ versus $\mathcal{A}_1$
is not linear either.
Instead, we use the sphere radius $R$ as a relevant length scale while
keeping $\mathcal{A}_1$ proportional to $R$ for self-similarity.
Thus, we grow a cluster until $\mathcal{A}_1$ reaches at $\phi_o R$ for 
various radius $R$, but with a fixed particle size, $\ld_o=1$, and
a fixed angle $\phi_o=\pi/2$. The angle $\phi_o$ is as such since
the circle (dashed lines in \Fig{cluster}) becomes a great circle
on a sphere.
If $N$ is the number of particles to fill the radius,
the fractal dimension is determined from $\ens{N}\sim R^{\Df}$.
From the statistics of 1000 clusters for each of $R$ in a geometrically 
increasing sequence from 79 to 400, we obtain the fractal dimensions,
$\Df \approx 1.70$, as shown in Table~\ref{tb:dim}.
\Fig{fracdim}(a) shows $\ens{N}$ versus $R$ in the three geometries.
The relative deviation of $\Df$ between geometries is surprisingly small
compared to the deviation of surface properties caused by the curvature.
For example, the area within the radius $(\pi/2)R$ on elliptic
(or hyperbolic) geometry is about 23\% smaller (or larger)
than the corresponding area on the Euclidean surface; however,
this factor only change the prefactor, not the exponent,
of the scaling, $\ens{N}\sim R^{\Df}$.

Our results suggest that $\Df$ is insensitive to the curvature because,
on small length scales comparable to the the particle size $\sqrt{\ld_0}$,
the surface is locally Euclidean.
This result is consistent with ADLA~\cite{adla,david05} whose fractal dimension is not 
affected by the background fluid flow. We conjecture that any conformally invariant transport-limited aggregation process on a $d$-dimensional curved manifold has same, universal fractal dimension as DLA in flat (Euclidean) space in $d$ dimensions.

More subtle statistics of the aggregates, however, are revealed by multifractal scalings. 
The multifractal properties~\cite{hentschel83,halsey86} related to from the 
probability measure, however, seem to depend on the curvature. Following 
Ref.~\cite{david99}, we measure the multifractal dimensions $D_q$ from the 
relation,
\begin{equation}
  \label{eq:Dq}
  \ens{\ld_N^q} \sim R^{-2qD_{2q+1}}.
\end{equation}
The averaging in \Eq{Dq} is made over $\ld_N$ at uniformly distributed 
angle. \Fig{fracdim}(b) shows the first three moments of 
$\ld_N$ as functions of $R$, and Table~~\ref{tb:dim} shows $D_q/\Df$ 
in the three geometries.
We note that $D_q$ for elliptic geometry are yet transient, influenced 
by finite-size effect. The multifractal dimensions do not satisfy the
inequality $D_q > D_{q'}$ for $q<q'$~\cite{hentschel83}.
Since the space itself is finite and the distance to {\it infinity}
is bounded, the instability in growth is pronounced in elliptic geometry.
The normalized center-of-charge fluctuation, 
$\ens{|\mathcal{A}_0|^2}^{1/2} / R$, is about 0.3 for on elliptic geometry 
although it decreases as $R$ increases.
For Euclidean and hyperbolic geometries, the fluctuations are
around 0.03 and 0.02 respectively.
It should also be noted that the growth probability on sphere is 
not harmonic since the far-field potential, $-\log|z|$, on plane
is mapped to the singular potential, $\log(\pi-\phi)$, around the north 
pole.

Even with the finite size effect, we conjecture that multifractal dimensions 
increase in the order of hyperbolic, Euclidean, and elliptic geometries.
The justification is that, when the measurements are similarly made for 
smaller latitudinal angles, $\phi_o = \pi/3$ and $\pi/6$, $D_q$ on elliptic 
and hyperbolic geometries converge to $D_q$ on Euclidean plane.
This also support that the slight difference in $D_q$ between Euclidean 
and hyperbolic geometries are not from statistical error.
We believe the dependence of $D_q$ on curvature is related to
the depth of fjords. On elliptic geometry, the circumference of a circle with 
radius $\phi_o R$ is $2\pi\sin\phi_o R$, shorter than $2\pi\phi_o R$ on 
Euclidean geometry. If we consider a pair of branches which
make the same opening angle on the two geometries,
the area between them is more screened on elliptic than on Euclidean
geometry. Therefore, we expect a different distribution of $\ld$.
Due to the competition on smaller circumference, tip-splitting events
are subdued, as observed in continuous growth~\cite{parisio01}, 
and eventually fewer branches survive.
The opposite argument applies to hyperbolic geometry, where the circumference
is given by $2\pi\sinh\phi_o R$, inter-branch area is less-screened, and 
tip-splitting is encouraged. In future work, it would be interesting to analyze the spectrum of $\ld$ 
and its connection to tip splitting events.

In summary, we have developed a mathematical theory of transport-limited growth on curved surfaces and applied it to DLA on two-dimensional surfaces of constant Gaussian curvature.
Our simulations suggest that the fractal dimension of DLA clusters is universal, independent of curvature, and depends only on the spatial dimension. The multifractal properties of DLA, however, seem to depend on curvature, since tip-splitting is related to the different degrees of screening the inter-branch areas. 
We conjecture that these results are hold in general, for any conformally invariant transport-limited growth process, in any number of dimensions.

We acknowledge helpful discussions with Ellak Somfai.

\end{document}